\documentclass[prl,twocolumn,superscriptaddress,longbibliography,showpacs,amsmath,amssymb]{revtex4-1}

\usepackage{soul}  

\usepackage{graphicx}
\usepackage{dcolumn}%
\usepackage{bm}

\usepackage[usenames,dvipsnames]{color}

\begin{document}

\title{Macroscopic polarization from antiferrodistortive cycloids in ferroelastic SrTiO$_3$}

\author{Andrea Schiaffino}
\affiliation{Institut de Ci\`encia de Materials de Barcelona
(ICMAB-CSIC), Campus UAB, 08193 Bellaterra, Spain}

\author{Massimiliano Stengel}
\affiliation{ICREA - Instituci\'o Catalana de Recerca i Estudis Avan\c{c}ats, 08010 Barcelona, Spain}
\affiliation{Institut de Ci\`encia de Materials de Barcelona
(ICMAB-CSIC), Campus UAB, 08193 Bellaterra, Spain}

\date{\today}
\begin{abstract}
Based on a first-principles based multiscale approach, we study the polarity ($P$) of 
ferroelastic twin walls in SrTiO$_3$. 
In addition to flexoelectricity, which was pointed out before, we identify
two new mechanisms that crucially contribute to $P$:
a direct ``rotopolar'' coupling to the gradients of the antiferrodistortive (AFD) oxygen tilts,
and a trilinear coupling that is mediated by the antiferroelectric displacement of the Ti atoms.
Remarkably, the rotopolar coupling presents a strong analogy to the mechanism 
that generates a spontaneous polarization in cycloidal magnets. 
We show how this similarity allows for a breakdown of macroscopic inversion symmetry 
(and therefore, a macroscopic polarization) in a periodic sequence of parallel twins.
These results open new avenues towards engineering pyroelectricity or 
piezoelectricity in nominally nonpolar ferroic materials.
\end{abstract}

\maketitle

 
  Domain walls (DW) in ferroic materials are a recognized source of 
  unusual physical effects, of practical interest for electronic 
  device applications.~\cite{Catalan12}
  Such properties are mainly due to the local modification of 
  the crystal structure, which can manifest itself via two distinct
  mechanisms.
  First, one or more components of the primary order parameter need to 
  change sign at the DW, and hence it will locally vanish. 
  This implies that latent secondary intabilities, suppressed in the 
  stable bulk phase because of the mutual competition between modes, 
  may in principle become active at the wall.
  Second, in the domain wall region one or more degrees of freedom undergo
  a large variation on a short length scale, which means that gradient couplings 
  (e.g., flexoelectricity) can have a strong impact on the physics.

  Among all different DWs, ferroelastic twin boundaries seem 
  very promising  candidates for emerging functionalities.
  Indeed, they locally break inversion symmetry, and hence can show a polar behaviour, 
  even if the bulk ferroelastic domains are nonpolar~\cite{Salje_review}.
  One of the most representative examples is CaTiO$_3$, 
  where the presence of a local polarity at
  the twin boundaries (TB) was first theoretically predicted via an empirical
  atomistic model~\cite{Artacho08}, and later experimentally confirmed~\cite{Salje12} by 
  transmission electron microscopy. 
  Interestingly, a recent first-principles~\cite{Picozzi14} 
  analysis has postulated an \emph{improper} origin of the polarization, which 
  would emerge from trilinear couplings between tilt modes that are enabled
  in the domain-wall region. 

  While polarity should, by symmetry, be present at the twin boundaries 
  of SrTiO$_3$ (STO) as well, the available supporting experimental evidence,
  based on the appearance of electromechanical resonance peaks in the low-temperature 
  regime~\cite{Scott13},
  is only indirect.
  Existing phenomenological works emphasize a flexoelectric origin of the 
  polarization at twin boundaries in SrTiO$_3$:
  A ferroelastic twin wall, by definition, separates two domains
  with different strain states; thus, a strain \emph{gradient} must 
  necessarily be present at the boundary. Since flexoelectricity 
  is a universal effect of all insulators, it must produce a net 
  polarization therein~\cite{Morozovska12N,Morozovska12}.  
  Such an interpretation was recently confirmed by numerical simulations based 
  on a simplified atomistic model~\cite{Salje16}, but a first-principles analysis 
  is still missing. Consequently, many fundamental and practical questions are 
  left, to date, unsettled; given the increasing importance of SrTiO$_3$ as
  a functional oxide material this is a timely moment for clarifying these matters.
 
  The first obvious question concerns the magnitude of the
  wall polarization: Phenomenological theories or empirical 
  potentials can hardly push their accuracy beyond order-of-magnitude 
  estimates, as many of the relevant coupling coefficients (e.g.
  the flexoelectric tensor) are difficult to access experimentally.
  The second, more profound, question concerns the very physical origin of 
  the couplings that may induce a polarization ($P$) at the wall: Is
  flexoelectricity really the end of the story in SrTiO$_3$? Or is
  there, similarly to the CaTiO$_3$ case,~\cite{Picozzi14} also an ``improper'' 
  contribution to $P$, directly related to the tilt pattern?
  And, if yes, is there a way to write such improper
  couplings as well-defined bulk properties of the material?

  To answer these questions with quantitative accuracy, a sound microscopic 
  analysis, e.g. as provided by density functional theory (DFT),
  is clearly mandatory. 
  Note that, in the past, DFT has been applied with success to a vast range of complex 
  ferroic systems, e.g. the so-called ``hybrid improper'' ferroelectrics~\cite{Benedek11}.
  Applying the same computational strategies (i.e., of inspecting the energy landscape 
  of the crystal in a vicinity of some high-symmetry reference structure)
  to a spatially inhomogeneous system such as a 
  domain wall, however, presents considerable methodological challenges.
  In particular, to correctly describe the local variation of the order 
  parameters near the wall one needs to consider \emph{gradient-mediated} couplings --
  these imply a breakdown of translational symmetry, and are therefore 
  difficult to calculate within DFT.
  As we shall illustrate shortly, recent advances in the first-principles theory 
  of flexoelectricity have  now opened the way towards overcoming such limitation.

  Here we present a novel multiscale theory of emergent
  polarity at improper ferroelastic walls, which relies on a 
  systematic perturbative approach to bridge the macroscopic
  and microscopic length scales.
  In particular, we 
  extract the relevant gradient-mediated terms in the
  Hamiltonian via a long-wavelength expansion of the linear and 
  nonlinear interatomic force constants (IFCs) of the 
  reference bulk phase. The calculated coupling coefficients
  are then incorporated in a one-dimensional continuum model,
  whose solutions yield the relaxed atomic structure of the 
  twin boundary.
  Application of this method to SrTiO$_3$ reveals that no less than
  three independent couplings contribute to the polar distortions at
  a twin wall: (i) the flexoelectric ``roto-flexo'' coupling  that was 
  already considered in earlier works~\cite{Gopalan11,Morozovska12N}; 
  (ii) a new ``rotopolar'' coupling,
  involving the tilt amplitude and the gradient of the tilt, which 
  generalizes and provides a formal basis to the ``improper'' mechanism of 
  Ref.~\onlinecite{Picozzi14}; (iii) a trilinear coupling mediated
  by the \emph{antiferroelectric} (AFE) displacement of the Ti atoms,
  which was never reported before, to the best of our knowledge.
  Our calculations show that contributions (i-iii) are comparable in
  magnitude, and they all must be included to gain a quantitative, or
  sometimes even qualitative, insight into the properties of a twinned
  SrTiO$_3$ sample. 
  Remarkably, the ``rotopolar'' coupling (ii) has the exact same form as
  that occurring in spiral magnets~\cite{Mostovoy06}; this implies that
  simple twinned structures can, in close analogy to spin cycloids, produce a 
  measurable macroscopic polarization in ferroelastic SrTiO$_3$. 
  We also find that twin walls in SrTiO$_3$ are much thicker (we
  find a characteristic length in excess of $5$ nm in
  the low-temperature limit) than previously thought,~\cite{CaoBarsch90} with a 
  reversed energy ordering between ``easy'' and ``hard'' types.~\cite{Morozovska12}

  \begin{figure}
    \begin{center}
    \includegraphics[width=3.3in]{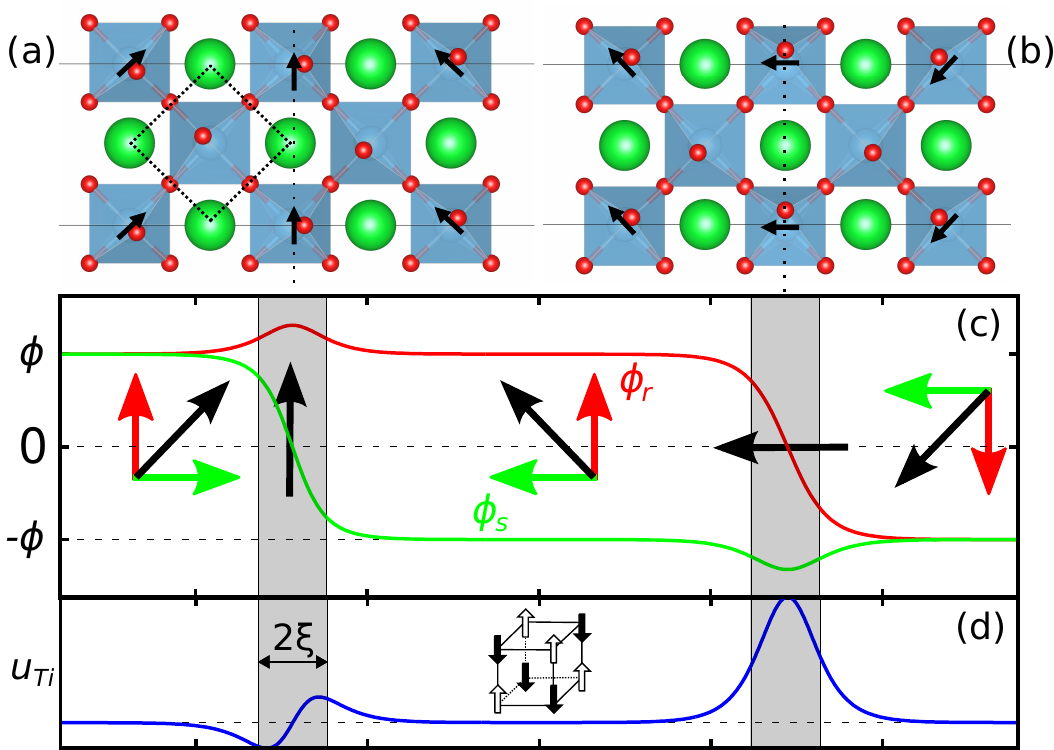}
    \end{center}
    \caption{\label{fig1} 
             (a-b): Schematic illustration (not to scale) of the two different types of 
             TBs considered in this work, respectively HH (a) and HT (b). Sr (large
             green balls), O (small red balls) and the oxygen octahedra are shown;
             dashed square indicates the primitive cell of the cubic reference phase;
             arrows indicate the local tilt vector.
             (c) Evolution of $\phi_s$ and $\phi_r$ across
             the two TBs. A local decomposition of the tilt vector (black arrows) into $s$ (green) and $r$ 
             (red) is also shown. The shaded area indicates the nominal wall thickness, $2\xi$. 
             (d) Amplitude of the $u_{Ti}$ mode in arbitrary units. The inset illustrates the AFE
             character of the Ti displacements, resembling spins in a G-type antiferromagnet.
             The length scale is in units of $\xi$.
             }
   \end{figure}

  Ferroelastic twins in SrTiO$_3$ occur between tetragonal domains (the
  tilt pattern is $a^0 a^0 c^-$ in Glazer notation) whose 
  respective AFD tilt axes are oriented at 90$^\circ$ with respect to each other.
  Following the usual procedure,~\cite{CaoBarsch90,Morozovska12} they can be conveniently represented 
  as a one-dimensional problem, where the relevant vector or tensor quantities are projected along 
  two directions that are either  perpendicular ($\hat s$) 
  or parallel  ($\hat r$) to the wall.
  (These correspond to the $[110]$ and $[1\bar 1 0]$ pseudocubic directions, respectively.)
  Depending on which component of the AFD pseudovector, $\phi_r$ or $\phi_s$, changes sign at the wall,
  we have two distinct types of twin boundaries; we shall indicate them as ``head-to-tail''
  (HT) and ``head-to-head'' (HH) henceforth (see Fig.~\ref{fig1}).

  With this in mind, we shall expand the energy around the reference cubic phase as follows ($i=r,s$; $\partial = \partial/\partial s$),
      \begin{equation}\label{energy}
     \begin{split}
    E =
     & \frac{C_{\alpha\beta}}{2}\varepsilon_{\alpha}\varepsilon_{\beta}  
            + \frac{\kappa}{2} |\bm{\phi}|^2 + A |\bm{\phi}|^4 + \frac{\chi^{-1}_0}{2} P^2 \\  
     &  - R_{i\alpha} \phi_i^2 \varepsilon_{\alpha}   -Q_{i}   \phi_{i}^2  P^2  
            - e_{\alpha} P^2 \varepsilon_{\alpha}  \\
     &  + \frac{D_i}{2} (\partial \phi_i)^2 + \frac{G}{2} (\partial P)^2 - f P \partial \varepsilon_{rs} 
            - W_{ij} P (\partial \phi_i) \phi_j \\
     &  + \frac{\kappa^{\rm Ti}}{2} (u^{\rm Ti})^2 + N P u^{\rm Ti} \phi_{s} + S (\partial \phi_r) u^{\rm Ti},
  \end{split}
   \end{equation}
   where the independent variables (all represented as continuous functions of the $s$ coordinate) are the 
   tilt amplitudes ($\phi_{r,s}$), the polarization ($P$) along $\hat r$ (we shall neglect
   the small antisymmetric components of ${\bf P}$ that are perpendicular to the wall plane~\cite{Picozzi14}),   
   which we associate with the amplitude of the ``soft'' zone-center optical 
   phonon, 
    the strain components in Voigt notation ($\varepsilon_{\alpha}$) and the amplitude of the antiferroelectric (AFE) 
   displacement of the Ti atoms ($u^{\rm Ti}$) along the direction that is orthogonal to both $\hat r$ and $\hat s$. (The
   motivation for including this latter degree of freedom will become clear shortly.)

   The first two lines of Eq.~(\ref{energy}) include the well-known 
   couplings that are active in the homogeneous crystalline phase, namely:
   the elastic energy, the double-well potential associated to the unstable 
   AFD modes ($\kappa <0$ and $A>0$), the susceptibility of the polar mode in
   the harmonic limit, and the rotostrictive, biquadratic and electrostrictive
   couplings involving $P$, $\phi_i$ and $\varepsilon_\alpha$.
  The third line contains gradient-mediated couplings, which
  are directly responsible for the properties of the domain walls.
  In particular, we have: the correlation term involving either the 
  AFD or the polar mode (these introduce an energy cost associated with spatial variations
  of the corresponding order parameter), the flexoelectric coupling
  between gradients of the shear strain and $P$, and the new 
  ``rotopolar'' term.~\footnote{
    This term bears some similarities to the ``flexoantiferrodistortive'' (FxAFD) coupling
    described in Ref.~\onlinecite{Morozovska13}. At difference with FxAFD, 
    however, the rotopolar coupling 
    allows for $W_{rs} \neq W_{sr}$, and 
    hence for a macroscopic polarization in a cycloidal structure.}
  Finally, the last line contains the terms involving the AFE
  amplitude: the harmonic restoring force of the
  mode, a trilinear term involving $P$ and $\phi_s$, and a 
  further harmonic term that couples $u^{\rm Ti}$ to the gradient of $\phi_r$.
  (In 3D, both $S$ and $N$ can be written as third-rank
  pseudotensors, $S\epsilon^{ijk}$ and $N\epsilon^{ijk}$, where $\epsilon^{ijk}$
  is the antisymmetric Levi-Civita symbol~\cite{Morozovska13}.)

  Note that $u^{\rm Ti}$ is associated to a zone-boundary ($R$-point) phonon, 
  i.e. it does not carry any polarization by itself; it is \emph{not}
  an active instability of the system either (it is actually quite ``hard'' --
  the calculated frequency is $\omega_{\rm Ti} \sim 420$ cm$^{-1}$).
  Yet, $u^{\rm Ti}$ strongly couples to spatial variations of the AFD modes, 
  which leads to two important  
  consequences: (i) it significantly lowers the cost of tilt gradients, 
  and hence the domain-wall energy (this effect can be
  understood as an effective rescaling of the coefficient $D_r$) and
  (ii) it contributes to the polarization via the trilinear coupling to
  $\phi_s$. 
  This clarifies, at a qualitative level, the physical motivation for its explicit 
  consideration; the quantitative impact is rather remarkable, too, as we 
  shall illustrate in the results section.

  We calculate the coupling coefficients of Eq.~(\ref{energy}) from first-principles,
  by using the local density approximation to DFT as implemented in the ABINIT~\cite{abinit} code~\cite{supp}.
  The homogeneous coupling coefficients are fitted to the energy landscape of the
  periodic crystal by distorting the lattice along the mode coordinates, 
  following the established practice.~\cite{improper,Benedek11}
  The harmonic gradient-mediated terms are extracted by loosely following the 
  strategy of Ref.~\onlinecite{Max16}, i.e. by first performing a long-wave expansion 
  of the dynamical matrix (calculated via linear-response theory~\cite{Baroni01}) around 
  either the $\Gamma$- or the R-point of the Brillouin zone, and by subsequently 
  performing the projections onto the relevant mode eigenvectors.
  
  The tricky part consists in the calculation of the \emph{nonlinear} gradient-mediated term,
  i.e. the rotopolar $W$-coefficients -- to the best of our knowledge, there are no 
  reported \emph{ab initio} calculations of such quantities (or, more generally, of such class of
  materials properties).
  We solve this technical obstacle by performing, as above, linear-response calculations of the 
  dynamical matrix as a function of the wavevector $q$.
  This time, however, we perform the phonon calculations on a 20-atom cell, where we 
  freeze in a small uniform tilt, $\phi_j$, by hand. 
  The two symmetry-allowed $W$-coefficients ($W_{rs}$ and $W_{sr}$) are then calculated by numerically 
  differentiating the dynamical matrix, $D$, with respect to \emph{both} parameters,
  \begin{equation}
  W_{ij} = \langle P | \, \frac{\partial^2 D(\phi_j,q)}{\partial \phi_j \partial q} \Big|_{\phi_j,q=0} |\phi_i \rangle.
  \end{equation}
  ($|P\rangle$ and $|\phi_i \rangle$ are mode eigenvectors.) 
  More extensive details on the technique will be reported in a forthcoming publication.
  
  Most of our calculated values are in good (sometimes excellent) agreement~\cite{supp} with
  earlier experimental and theoretical estimates,~\cite{UweSakudo76,SaiVanderbilt00} whenever the latter are available. There are two
  important exceptions, though, which deserve a separate discussion. 
  The harmonic coefficients that are linked to the polarization and AFD
  degrees of freedom, respectively $\chi_0^{-1}$ and $\kappa$, are notoriously difficult
  to capture within standard approximations to DFT; as a consequence, the spontaneous tilt
  angle at zero temperature is severely overestimated ($5-6^\circ$ versus the experimental 
  value of $2.1^\circ$), 
  and the reported DFT values for the static 
  dielectric constant of SrTiO$_3$ are remarkably erratic.
  Luckily, these two coefficients are by far the easiest to estimate from the experimental
  data (tilt angle and dielectric susceptibility as a function of temperature, respectively);
  for this reason, we replace their first-principles values with a phenomenological Curie-Weiss
  function of $T$~\cite{supp}.
  Other coefficients in our model are expected to be quantitatively accurate
  and to depend only weakly on temperature.

     \begin{figure}
    \centering
    \includegraphics[width=2in]{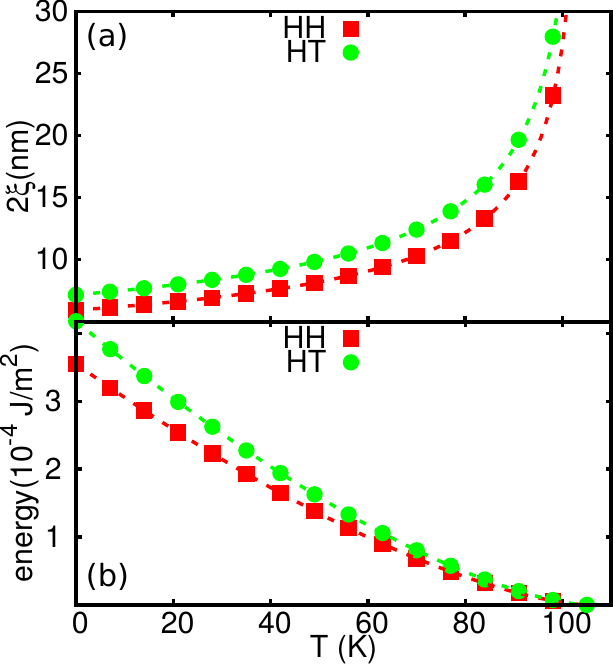}
    \caption{\label{fig2} 
             (a) Domain wall thickness, $2\xi$, as function of temperature.
             (b) Formation energy per surface unit as function of temperature.}
   \end{figure}

  To calculate the two stable domain wall structures by means of Eq.~(\ref{energy}),
  we implement the continuum equations on a discrete 1D mesh, and express the gradient terms
  as nearest-neighbor interactions; energy minimization proceeds via a steepest-descent algorithm.~\cite{supp}
  In Fig.~\ref{fig1} we plot the equilibrium configuration of the primary AFD order parameters
  at the two types of domain walls.
  The antisymmetric (with respect to $s \rightarrow -s$) AFD component ($\phi_{\rm A}$)
  can be very well fitted with a kink-type solution,~\cite{Salje_review,CaoBarsch90} $\phi_{\rm A}(s) \sim \tanh(s/\xi)$, 
  while the symmetric component shows a characteristic bump in correspondence of the boundary.
  (The squared modulus of the tilt pseudovector is roughly preserved throughout the
  structure.)
  Remarkably, we predict (Fig.~\ref{fig2}) that the wall widths are 
  $2\xi \sim 7-8$ nm, i.e. almost one order of magnitude thicker than the 
  established literature values for either SrTiO$_3$~\cite{CaoBarsch90,Morozovska12} 
  or other ferroelastic materials.~\cite{Salje_review}
  This discrepancy can be traced back to our calculated $D_{i}$ 
  gradient coefficients, which are much
  larger than the commonly used literature values~\cite{CaoBarsch90}. 
  That the domain 
  walls are so thick fits with the experimental observation that they
  are highly mobile, even at the lowest temperatures~\cite{Salje000}.
  (In retrospect, this result also justifies the continuum approximation that we 
  use, which is expected to be accurate at these length scales.)
  Another surprise comes from the energetics: HT walls, which were
  formerly believed~\cite{Morozovska12} to be the ``easy'' type of twin boundary, 
  are, in fact, slightly more costly than HH walls.
  Note that the expected~\cite{CaoBarsch90} scaling of the thickness, $2\xi$, 
  and energy, $E$, as a function of $\kappa$ and the $D_i$ coefficient are 
  accurately respected by our results: $\xi_i \sim \sqrt{D_i /\kappa} $
  and $E_i \sim \kappa^{2} \sqrt{D_i /\kappa} $,

  \begin{figure}
    \centering
    \includegraphics[width=3.3in]{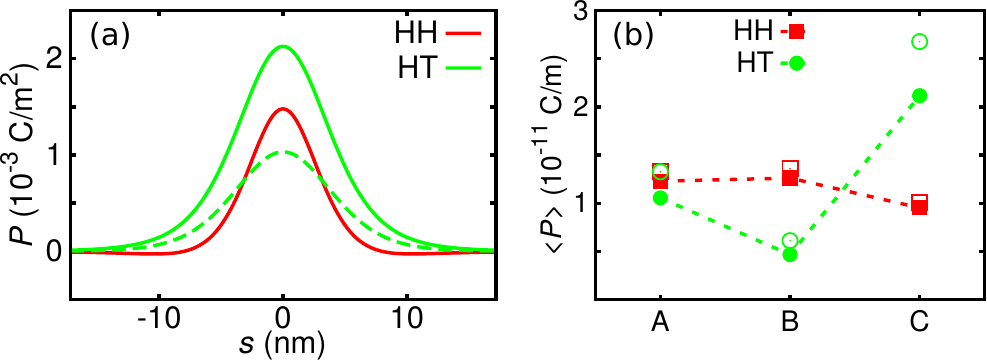}
    \caption{\label{fig3} 
             (a): Polarization profile across the two DWs; the dashed
             line refers to the result (B), i.e. without including 
             $u^{\rm Ti}$ in the simulation.
             (b): Total polarization integrated across the
             DW as function of the addition of different couplings
             in the Hamiltonian. Empty and filled symbols refer to 
             the results obtained while excluding or including the
             biquadratic and electrostrictive terms. The polarization
             vector is always oriented towards the apex of the twin boundary. }
   \end{figure}

 We move now to the main result of this work, regarding the 
 induced electrical polarization at either type of domain boundary.
  To present our findings, we shall work at a  reference temperature of $T_{\rm ref} =80$ K,
  where the associated tilt is $1.4^\circ$~\cite{CaoBarsch90}.
  ($T_{\rm ref}$ is chosen out of convenience, as it is precisely the
  temperature at which the calculated value of $\chi_0$ matches the 
  experimentally measured~\cite{Weaver59} dielectric constant of SrTiO$_3$;
  the main conclusions that will be
  presented in the following are, nevertheless,
  valid at any temperature below $T_\phi$=105 K, where $T_\phi$ is 
  the ferroelastic transition temperature.)

  Looking back at the energy, Eq.~(\ref{energy}), one can see that there are three
  \emph{improper} (linear in $P$) mechanisms that can induce a 
  polarization at the wall: flexoelectricity, rotopolar copuling, and the trilinear
  coupling mediated by $u^{\rm Ti}$. 
  To quantitatively separate their individual role, we shall start with a simpler
  Hamiltonian where we artificially set $W_{ij}=S=N=0$ 
  (i.e., a polarization can only be induced via flexoelectricity), and progressively 
  switch on the new couplings, while monitoring their impact on the total 
  (integrated) polarization, $\langle P \rangle_{\rm HH,HT}$, at either wall.
  Note that the electrostriction and biquadratic couplings also have an
  impact on $\langle P \rangle$, although they fall in a different 
  category (they both go like $P^2$); to quantify their importance, we shall 
  perform our computational experiment twice, either with or without the latter two terms.

  From our modified Hamiltonian, we obtain (Fig.~\ref{fig3}, A points) that
  the polarity of HH and HT walls is essentially the same. (The small difference
  is only due to the $P^2$ terms.) This is easily understood: Flexoelectricity is 
  the only mechanism at play here, and therefore the ``geometric'' field acting on
  $\langle P \rangle$ can only depend~\cite{Salje16} on the total discontinuity in the 
  shear strain component, $\Delta \varepsilon_{rs}$. Of course, $\Delta \varepsilon_{rs}$ 
  is the same at both types of walls.

  Next, we switch on the new rotopolar coupling (B).
  As we anticipated in the introductory paragraphs,
  this mechanism clearly distinguishes one type of domain wall from the other:
  Its contribution to $\langle P \rangle$ is dominated by $W_{rs}$
  at HT walls, and by $W_{sr}$ at the HH walls. $W_{sr}$ is very small~\cite{supp},
  so at HH walls the polarization is almost unaffected; conversely, 
  $W_{rs}$ is large and almost cancels the flexoelectric effect at 
  HT walls.
  Thus, the rotopolar term has a central physical importance: it 
  provides us with the possibility of ``engineering'' a macroscopic 
  ferrielectric-like polarization, $P_{\rm mac}$, in a periodic twin wall structure. 
  It suffices to alternate HT and HH wall types, as illustrated in Fig.~\ref{fig1},
  to obtain ($L$ is the average domain width) 
  $P_{\rm mac} = (\langle P \rangle_{\rm HH} - \langle P \rangle_{\rm HT}) / (2L) \neq 0$.
  (In absence of this term, $P_{\rm mac}$ would vanish by symmetry
  in a sequence of parallel twins.~\cite{Salje16})
  Interestingly, even without doing the calculations there is an insightful 
  visual proof that such a structure indeed does break macroscopic 
  inversion symmetry.
  By following the evolution of the AFD pseudovector across the structure,
  one can easily identify a counterclockwise rotation of $\bm{\phi}$ for
  increasing $s$. 
  Such a pattern, in strong analogy to a spin cycloid,~\cite{Mostovoy06}
  does not posses inversion symmetry along ${\hat r}$, which proves our point.

  If we stopped our analysis here, we would be forced to conclude that 
  $\langle P \rangle_{\rm HH} > \langle P \rangle_{\rm HT}$.
  However, as we show in Fig.~\ref{fig3}, the trilinear coupling between $u^{\rm Ti}$,
  $P$ and the AFD tilts has a dramatic impact on $\langle P \rangle$, 
  to the point that it reverses (C) the ordering of $\langle P \rangle_{\rm HH}$ 
  and $\langle P \rangle_{\rm HT}$ (and hence the sign of $P_{\rm mac}$).
  As expected, the most affected wall type is the HT, where 
  $\phi_r$ changes sign. (The 
  trilinear coupling can be thought as an effective additional
  contribution to $W_{rs}$, while its contribution to $W_{sr}$ 
  vanishes; note also in Fig.~\ref{fig1} the much larger amplitude 
  of $u^{\rm Ti}$ at the HT wall.)
  Note that, in all cases, the effect of the biquadratic/electrostrictive 
  couplings is a systematic suppression, somewhat stronger at the HT walls, 
  of the gradient-induced polarization (compare empty and filled symbols in
  Fig.~\ref{fig3}).
  This observation implies that the $P^2$ terms alone are unlikely to trigger
  a ferroelectric state at either type of twin boundary, and corroborates 
  improper mechanisms as the main driving force for $P$.

  These results open new perspectives for breaking macroscopic
  inversion symmetry (and hence engineering an effective piezoelectric and/or pyroelectric 
  behavior) via twinning -- Ref.~\onlinecite{Salje16} explored the potential of 
  defects (kinks, junctions, vortices) in the domain wall topology, while here we demonstrate
  that a macroscopic $P$ can emerge even in ``ideal'' ferroelastic structures.
  These arguments can be readily generalized to other materials systems: For example,
  the improper mechanism pointed out~\cite{Picozzi14} at CaTiO$_3$ twins can be simply 
  (and quantitatively) rationalized as a rotopolar coupling.
  More generally, our work open new avenues for materials design via 
  domain wall engineering, an increasingly popular strategy 
  where new functionalities emerge from spatial inhomogeneities, rather than 
  the uniform crystalline phase itself.

  \begin{acknowledgments}
We acknowledge the support of MINECO-Spain through Grants No. FIS2013-48668-C2-2-P,
MAT2016-77100-C2-2-P and SEV-2015-0496, and of Generalitat de Catalunya (Grant No. 2014 SGR301).
This project has received funding from the European Research Council (ERC) under the European 
Union's Horizon 2020 research and innovation programme (grant agreement No. 724529).
Calculations were performed at Supercomputing Center of Galicia (CESGA).
  \end{acknowledgments}


\bibliography{text}

  \begin{widetext}
  \newpage

\begin{center}

{\large\bf
Supplementary notes: Computational Details}

\end{center}

   \begin{table*}[!b]
   \begin{center}
   \begin{tabular}{cr@{.}lr@{.}lr@{.}lcrl}
   \hline 
   \hline 
       &  \multicolumn{2}{c}{US}  & \multicolumn{2}{c}{SV} &  \multicolumn{2}{c}{this work}   & \multicolumn{3}{c}{a.u.}  \\
   \hline   
   $\kappa$ & $-$3&01  & $-$22&5 & \textit{$-$20}&\textit{62} & 10$^{-6}$  & Ha & bohr$^{-5}$ \\ 
   
   $A$ & 5&16  & 4&92 & 5&26 & 10$^{-5}$ & Ha & bohr$^{-7}$ \\ 
    
   $C_{11}$ &  11&43  & 13&02 & 13&14  & 10$^{-3}$ & Ha & bohr$^{-3}$ \\
   
   $C_{12}$ & 3&64  & 3&30     & 3&83  & 10$^{-3}$ & Ha  & bohr$^{-3}$ \\ 
   
   $C_{44}$ & 4&32  & \multicolumn{2}{c}{}        & 4&16   & 10$^{-3}$ & Ha & bohr$^{-3}$ \\ 
     
   $R_{11}$ & 1&23  & 1&68     & 1&95 & 10$^{-4}$ & Ha & bohr$^{-5}$ \\ 
   
   $R_{12}$ & $-$2&37  & $-$2&70 &$-$2&74 &  10$^{-4}$ & Ha & bohr$^{-5}$ \\  
   
   $R_{44}$ & $-$2&18  & \multicolumn{2}{c}{} &    $-$2&42 &  10$^{-4}$ & Ha & bohr$^{-5}$ \\  
 
   $Q_{r}$ & \multicolumn{2}{c}{}  & \multicolumn{2}{c}{} &   $-$0&28  & 10$^{-1}$ & Ha & bohr$^{-1}$ \\  
   
   $Q_{s}$ & \multicolumn{2}{c}{}  & \multicolumn{2}{c}{} &   $-$1&95  & & Ha & bohr$^{-1}$ \\   

   $N$ &     \multicolumn{2}{c}{} & \multicolumn{2}{c}{} & $-$1&53 &  10$^{-2}$ & Ha & bohr$^{-3}$ \\ 

   $\chi_0$  & \multicolumn{2}{c}{}     & \multicolumn{2}{c}{}  &  120&00  &  &Ha&bohr  \\
  
   $\kappa^{\rm Ti}$   & \multicolumn{2}{c}{}   &  \multicolumn{2}{c}{} &   3&54 & 10$^{-3}$& Ha&bohr$^{-5}$ \\
  
   $D_{r}$ & \multicolumn{2}{c}{}  & \multicolumn{2}{c}{} & 1&95   & 10$^{-3}$& Ha& bohr$^{-3}$ \\ 

   $D_{s}$ & \multicolumn{2}{c}{}  & \multicolumn{2}{c}{} & 1&00  & 10$^{-3}$ &Ha& bohr$^{-3}$ \\ 

   $S$ & \multicolumn{2}{c}{}  & \multicolumn{2}{c}{} & 1&29 &  10$^{-3}$& Ha& bohr$^{-4}$  \\

   $e_x$  & \multicolumn{2}{c}{}   & \multicolumn{2}{c}{} &  $-$0&18 & &Ha &bohr \\
    
   $e_s$  & \multicolumn{2}{c}{}   & \multicolumn{2}{c}{} &     1&31 & &Ha &bohr \\
   
   $e_r$  & \multicolumn{2}{c}{}   & \multicolumn{2}{c}{} &     1&84 & &Ha &bohr \\

   $G$ & \multicolumn{2}{c}{}  & \multicolumn{2}{c}{} & 5&43 & &Ha& bohr$^3$\\
   
   $f$ & \multicolumn{2}{c}{}  & \multicolumn{2}{c}{} &  $-$4&70 &  10$^{-2}$ &Ha&\\ 

   $W_{rs}$ & \multicolumn{2}{c}{}  & \multicolumn{2}{c}{} & 2&11  &  10$^{-3}$& Ha& bohr$^{-2}$ \\ 

   $W_{sr}$ & \multicolumn{2}{c}{}  & \multicolumn{2}{c}{} & 0&29  &  10$^{-3}$&Ha&bohr$^{-2}$ \\

   \hline
   \hline  
   \end{tabular}
   \end{center}
   \caption{ Calculated model parameters compared with the available literature data
            (US=Uwe and Sakudo [20], SV=Sai and Vanderbilt [21]). 
            The calculated value of $\kappa$ is reported in italics, 
            as we replaced it with a phenomenological function of temperature,
            $\kappa = \alpha_0 (T - T_\varphi)$. 
            (We used $T_\varphi$ = 105 K, $\alpha_0 = -0.013$ Ha/(bohr$^5$ K).) 
            At 80 K, $\varphi$ = 1.4$^\circ$, consistent with the experimental value. 
            [11] Note that at 80 K, the measured dielectric constant is approximately 
            consistent with our zero-temperature first-principles calculations, $4\pi \chi_0 \sim$ 1500.  } 
   \end{table*}


  \subparagraph*{Ab initio calculations}
  The \textit{ab initio} calculations have been performed with 
  norm-conserving pseudopotentials, taking into account explicitly 10 electrons for Sr, 
  12 for Ti and 6 for O. 
  The pseudopotentials have been generated using the FHI98PP code
  and the exchange-correlation term has been treated using
  the local-density approximation (LDA).
  Finally the energy cut-off used is 70 Ha and in all the first principle simulations
  we have employed a Monkhorst-Pack mesh equivalent to 8x8x8 grid in the primitive cubic
  cell.

  With these input parameters we get a bulk lattice constant value 
  for the cubic structure of $7.2675$ bohr.
  The phonon dispersion along the directions $\Gamma$-M and R-X 
  for this same reference structure is shown in fig.~4.
  As it is possible to see, we get a polar soft mode which is not
  unstable. 
  The only instability present in our structure is at R, and it is
  associated to the AFD of the oxygen atoms.

  \subparagraph*{1D model}
  
  Both the two types of ferroelastic TBs discussed in the
  main text have been studied employing a 1D model.
  This is possible because, by symmetry, it easy to see
  that the polar component along $\hat{s}$ is antisymmetric
  respect to the DW and thus does not contribute to the
  total polarization.
  Then the only polar component relevant for a macroscopic
  polarization at the DW is $P_r$.
  Moreover the translational symmetry is preserved along 
  both $\hat{r}$ and $\hat{x}$, which is the direction perpendicular
  to both $\hat{r}$ and $\hat{s}$.

  As consequence of these symmetries the following mechanical 
  boundary conditions must be imposed: $\varepsilon_r$ and $\varepsilon_x$
  are fixed to their correspondent value of the bulk AFD phase.
  Instead $\varepsilon_s$ and $\varepsilon_{rs}$
  are free to relax during the energy minimization 
  in order accomodate the deformation due to the ferroelastic twin walls. 
  A profile of the strain components as function of $\hat{s}$ is given in fig.~5.
  
  The 1D problem of minimizing eq.~1 of the main text has been
  solved adopting the steepest descent algorithm.
  The mesh used to discretize the previous equation
  has the atomic resolution,
  i.e. each point of the mesh corresponds to an atomic layer
  along the direction $\hat{s}$.
  The gradient terms have been calculated on the 1D mesh 
  using a  symmetric nearest-neighbor formula.
  Finally we have impose periodic boundary conditions
  to our 1D problem.
  As consequence of this choice in our simulations of ferroelastic domain walls,
  we have always to take into account at least two opposite walls.

  \begin{figure}
  \begin{center}
   \includegraphics[width=3.5in]{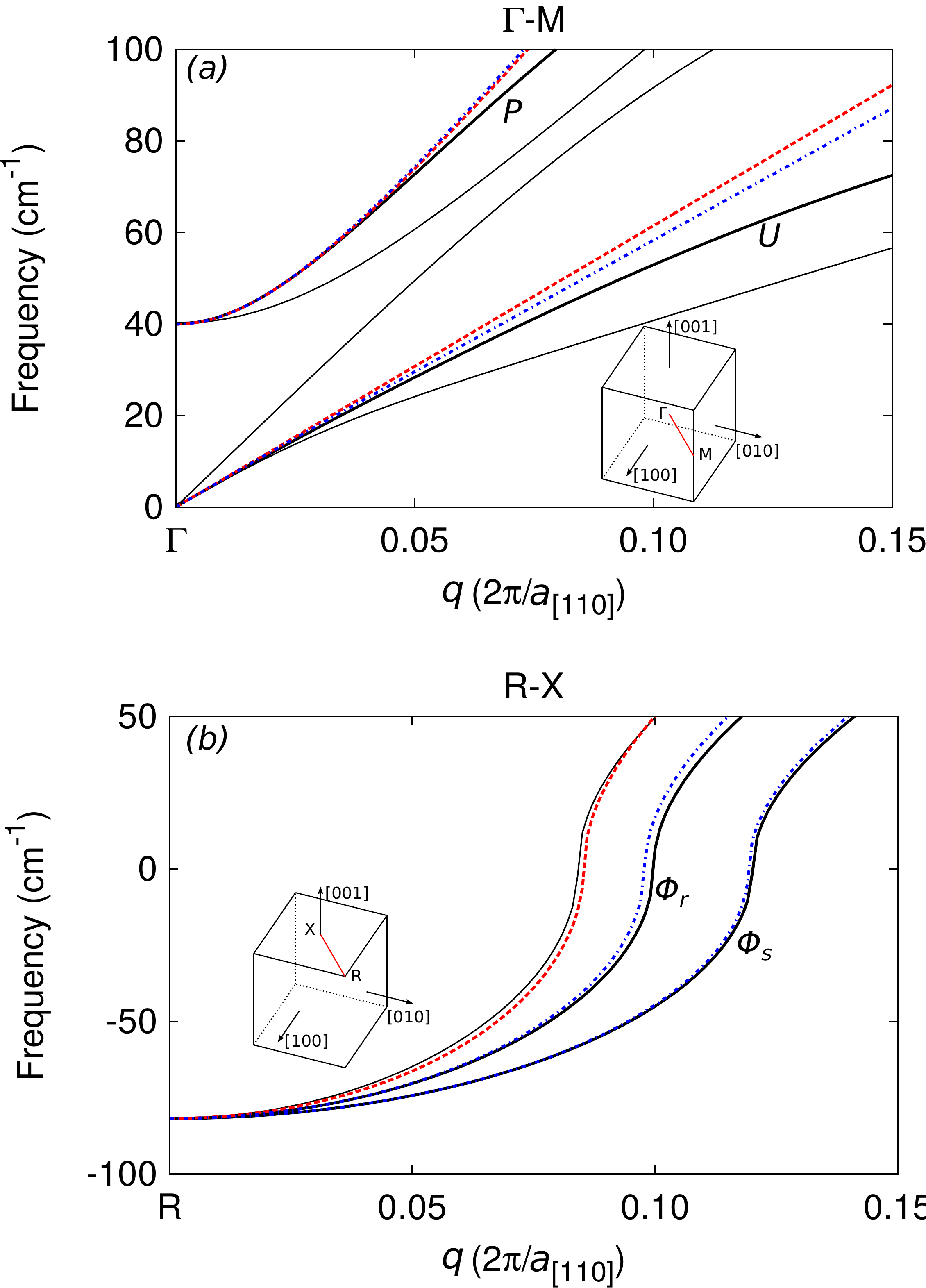}
   \end{center}
   \caption{\label{fig1} 
              Here we shown the phonon dispersion as function
              of the wavenumber $q$ in reduced coordinates,
              where $a_{[110]}=a_0\sqrt{2}$
              and $a_0$ is the cubic lattice parameter of STO.
              In figure (a) the phonon dispersions are along
              the $\Gamma$-M direction (as shown by the inset) 
              and they are calculated on one side
              using the Fourier interpolation of a 3D q-mesh (black-solid curves) 
              and on the other side
              using the two dimensional space defined by the transversal
              acoustic and center-zone soft polar mode
              (red dashed and blue dot-dashed), which approximate 
              the ticker black branches, $U_r$ and $P_s$, around $\Gamma$.
              In particular: the red dashed lines show the phonon bands 
              calculated using only the self-correlation of the
              acoustic and soft polar mode, which are respectively 
              the elastic constant and the $G$ coupling;
              the blue dot-dashed lines are calculated taking into account
              also the fexoelectric interaction term. 
              In a similar way figure (b) shows the phonon dispersion 
              around R, in direction [110], of the unstable AFD modes in cubic STO.
              The labels identify the branch associate to the oxygen octahedron
              rotation along $\hat{s}$, $\phi_s$, and along $\hat{r}$, $\phi_r$,
              while the third branch is the dispersion of the AFD with rotational
              axis perpendicular to both $\hat{s}$ and $\hat{r}$. 
              The red dashed line is calculated using only the AFD self dispersion,
              $D$, of $\phi_r$ while the blue dot-dashed line takes into account also the
              trilinear interaction term, $S$.
              }
  \end{figure}

  \begin{figure}
  \begin{center}
   \includegraphics[width=3.5in]{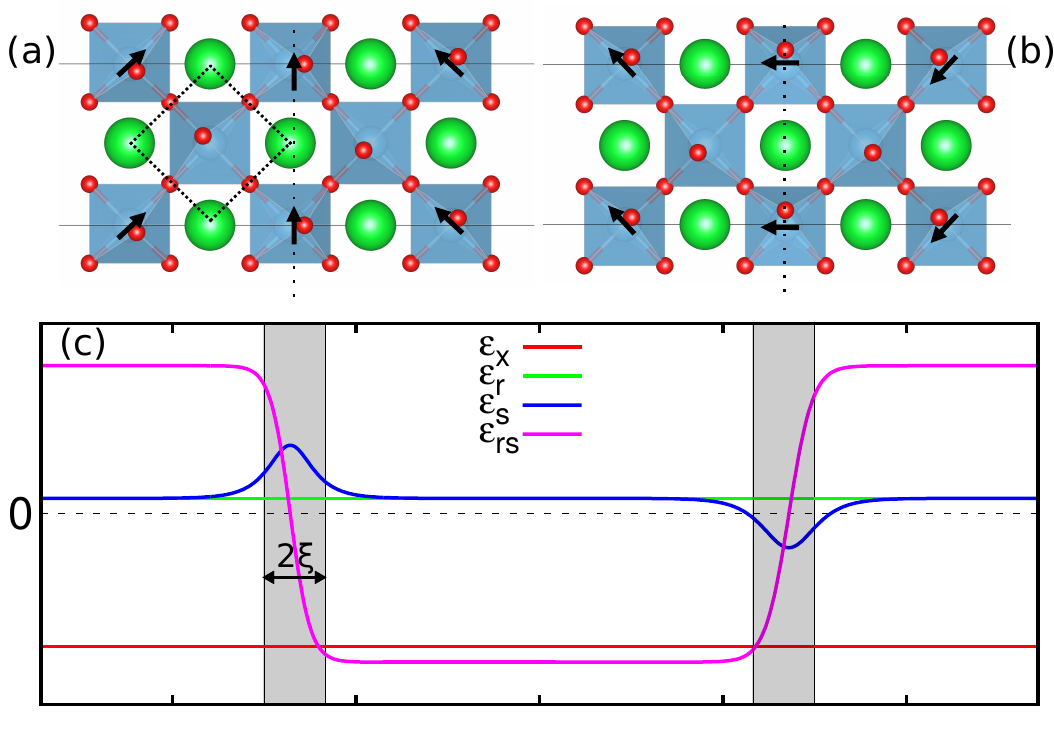}
   \end{center}
   \caption{\label{fig2} 
               Cartoons (a) and (b) are sketches of the 
               two ferroelastic twin walls studied in this work, respectively
               the HH and HT walls.
               In (c) are shown the strain components across the two DWs.
               $\varepsilon_r$ and $\varepsilon_x$ do not change across the walls because 
               they are kept fix to their bulk AFD value, in order to satisfy the mechanical 
               boundary conditions for a periodic system.
               Note that the cartoon (a) and (b) are on top (c) just as reference
               but they do not have the same scale.
              }
  \end{figure}

  \end{widetext}

\end{document}